\documentclass[a4paper,preprint,aps,prd,eqsecnum,showpacs,nofootinbib]{revtex4}
\usepackage[T1]{fontenc}
\usepackage[latin1]{inputenc}
\usepackage{amsmath}
\usepackage{graphics}
\usepackage{amssymb}

\makeatletter

\newcommand{\sla}[1]{\rlap{$#1$}/}

\newcommand{\comment}[1]{}

\makeatother

\begin{document}

\newcommand{\msb}{\overline{\mathrm{MS}}}

\newcommand{\tr}[1]{\mathrm{Tr}\left\{ #1 \right\} }


\pacs{11.10.Kk,11.10.Hi,12.10.Dm,12.10.Kt}

\preprint{
\begin{tabular}{r}
FTUV--03--0211\\ 
IFIC/03--04 
\end{tabular}
}

\title{
Can power corrections be reliably computed \\ 
in models with extra dimensions? }

\author{J.F. Oliver}

\affiliation{Departament de Física Teòrica and IFIC, Universitat de València -CSIC\\
 Dr. Moliner 50, E-46100 Burjassot (València), Spain}

\author{J. Papavassiliou}

\affiliation{Departament de Física Teòrica and IFIC, Universitat de València -CSIC\\
 Dr. Moliner 50, E-46100 Burjassot (València), Spain}

\author{A. Santamaria}

\affiliation{Departament de Física Teòrica and IFIC, Universitat de València -CSIC\\
 Dr. Moliner 50, E-46100 Burjassot (València), Spain}

\begin{abstract}
We critically revisit the issue of power-law running
in models with extra dimensions.
The analysis is carried out in the context of a higher-dimensional
extension of QED, with the extra dimensions compactified on a torus.
It is shown that a naive $\beta$ function, which simply
counts the number of modes, depends crucially on the way the thresholds of the
Kaluza-Klein modes are crossed.
To solve these ambiguities we turn to the vacuum polarization, which,
due to its special unitarity properties,  guarantees the physical
decoupling of the heavy modes. This latter quantity, calculated in the context of dimensional
regularization, is used for connecting the low energy gauge coupling with 
the coupling of the $D$-dimensional effective field theory. 
We find that the resulting relation contains only logarithms of the relevant
scales, and no power corrections. 
If, instead, hard cutoffs  are used to regularize the theory, 
one  finds power corrections, which  could  be interpreted  as  an  additional
matching between  the effective higher-dimensional model and
some unknown,  
more complete  theory.  The  possibility of
estimating  this  matching  is examined in the context of a 
toy model. The  general
conclusion is  that, in the  absence of any additional  
physical principle, the
power corrections depend strongly on the details
of the underlying theory.
Possible consequences of this analysis for
gauge coupling unification in  theories with extra dimensions are 
briefly discussed.

\end{abstract}
\maketitle

\section{Introduction}

The study of models with extra dimensions has received a great deal of
attention recently~\cite{Arkani-Hamed:1998nn,Arkani-Hamed:1998rs,Antoniadis:1990ew,
Antoniadis:1994jp}, mainly because  of the plethora of theoretical and
phenomenological ideas associated with  them, and the flexibility they
offer  for  realizing   new,  previously  impossible,  field-theoretic
constructions. One of the  most characteristic features of such models
is  that  of  the  {}``early  unification{}'': the  running  of  gauge
couplings is  supposed to be modified  so strongly by  the presence of
the tower of  KK modes, that instead of  logarithmic it becomes linear,
quadratic,    etc,    depending     on    the    number    of    extra
dimensions~\cite{Bachas:1998kr,Dienes:1998vh,Dienes:1998vg,Ghilencea:1998st,
Carone:1999cb,Delgado:1999ba,Floratos:1999bv,Frampton:1999ue,Kakushadze:1998vr,Kakushadze:1999bb,
Perez-Lorenzana:1999qb,Dumitru:1999ji,Masip:2000yw,Arkani-Hamed:2001vr,
Berezhiani:2001ub,Hebecker:2002vm,Kubo:1999ua}. 
Specifically, it has  been widely argued 
that the gauge couplings
run as \( \mu ^{\delta } \),  where the \( \delta \) is the number of
compact  extra   dimensions.  Thus,   if  the  extra   dimensions  are
sufficiently large, such  a behavior of the couplings  could allow for
their unification at  accessible energies, of the order  of a few TeV,
clearly an exciting possibility.

The assertion  that gauge-couplings display power-law  running is based
on rather intuitive arguments: In \(  \msb \) schemes the QED \( \beta
\) function  is proportional to the number  of {}``active{}'' flavors,
namely  the  number  of  particles lighter  than  the  renormalization
scale.  Using this  argument, and  just counting  the number  of modes
lighter than \( \mu \), one easily finds that the ``$\beta$ function''
of QED  in models  with extra  dimensions grows as  \( \mu  ^{\delta }
\). This  behavior is also  justified by explicit calculations  of the
vacuum polarization of the photon using hard cutoffs; since the cutoff
cannot be removed, due to  the non-renormalizability of the theory, it
is  finally identified  with  the renormalization  scale, a  procedure
which eventually  leads to a similar  
conclusion~\cite{Dienes:1998vh,Dienes:1998vg} (but  with the final
coefficient  adjusted by  hand  in order  to  match the  naive
expectation in \( \msb \)).

Even  though these arguments  are plausible,  the importance  of their
consequences   requires   that  they   should   be  scrutinized   more
carefully~\cite{Contino:2001si}.  In particular, the argument based on
\( \msb \) running is rather tricky.  As it is well known, the \( \msb
\)  scheme,  because  of  its  mass  independence,  does  not  satisfy
decoupling,  already  at   the  level  of  four-dimensional  theories.
Instead, decoupling  has to be  \textit{imposed} by hand every  time a
threshold  is  passed:  one  builds  an  effective  theory  below  the
threshold,  \(  m  \),  and   matches  it  to  the  theory  above  the
threshold.  This  matching  is  carried  out by  requiring  that  some
physical amplitude or Green's  function (i.e. the effective charge) is
the  same  when calculated  using  either  theory,  at energies  where
\textit{both} theories are reliable, namely  at \( Q^{2} \) much below
the threshold.  Then, since the  renormalization scale, \( \mu  \), is
still a free parameter, one chooses \( \mu \) around \( m \), in order
to avoid  large logarithms in the  matching equations. In  the case of
gauge    couplings    and    \(    \msb    \)    schemes    with    \(
\tr{I_{\mathrm{Dirac}}}=4  \)  one  finds  (at one  loop)  that  gauge
couplings are continuous at \(  \mu =m \). This statement is, however,
extremely     scheme     dependent:     just    by     choosing     \(
\tr{I_{\mathrm{Dirac}}}=2^{D/2}  \) it  gets completely 
modified (see for instance \cite{Rodrigo:1993hc}) .  In
addition to  these standard ambiguities, a new  complication arises in
the   context  of  higher-dimensional   models.  In   particular,  the
aforementioned procedure requires that  the different scales be widely
separated in  order  to   avoid  that  higher  dimension  operators,
generated in the process  of matching, become important.  However, the
condition  of having  well-separated thresholds  is  rather marginally
satisfied  in the  case  of an  infinite  tower of  KK  modes with  \(
M_{n}=nM_{c} \) (\( M_{c} \)  is the compactification scale). In fact,
as we will see in detail  later, the results obtained for a \( \beta
\) function  that just counts the number of active modes depend very strongly  
on the prescription chosen  for the
way the various thresholds are crossed.

As has  been hinted above,  the deeper reason behind  these additional
type of ambiguities is the fact that, gauge theories in more than \( 4
\)  dimensions, compactified or  not, are  not renormalizable.  At the
level of the 4-dimensional theory  with an infinite number of KK modes
the non-renormalizability manifests itself  by the appearance of extra
divergences,  encountered when  summing  over all  the  modes. If  the
theory  is not  compactified  the non-renormalizability  is even  more
evident, since  gauge couplings  in theories with  \( \delta  \) extra
dimensions  have dimension  \( 1/M^{\delta  /2} \).   Therefore, gauge
theories  in extra  dimensions should  be treated  as  effective field
theories   (EFT).  Working   with  such   theories   presents  several
difficulties, but, as we have learned  in recent years, they can also
be  very  useful. In  the  case of  quantum  field  theories in  extra
dimensions,  there is  no alternative:  basic questions,  such  as the
calculation of  observables or the unification of  couplings, can only
be addressed in the framework of the EFT's. However, before attempting
to answer  specific questions related  to the running of  couplings in
the extra-dimensional  theories, one should first clarify  the type of
EFT one is going to use, since  there are, at least, two types of
EFT~\cite{Georgi:1993qn}:
In one  type, known  as ``Wilsonian EFT''  
(WEFT)~\cite{Wilson:1974jj}, 
one  keeps only
momenta  below some  scale \(  \Lambda \),  while all  the  effects of
higher momenta or heavy particles  are encoded in the couplings of the
effective  theory.  This  method  is  very  intuitive  and  leads,  by
definition, to finite  results at each step; however,  the presence of
the cutoff in all expressions  makes the method cumbersome to use, and
in the particular  case of gauge theories difficult  to reconcile with
gauge-invariance.  The  WEFT approach has already been  applied to the
problem  of  running  of  couplings  in theories  with  compact  extra
dimensions,   but    only   for   the   case    of   scalar   theories
\cite{Kubo:1999ua}. Within  the context of another type  of EFT, often
termed ``continuum  effective field theories''  (CEFT) (see for instance
\cite{Weinberg:1979kz,Weinberg:1980wa,Hall:1981kf,Georgi:1993qn,
Leutwyler:1994iq,Manohar:1996cq,Pich:1998xt}),
one allows
the momenta of  particles to vary up to  infinity, but heavy particles
are removed from the spectrum at low energies. As in the WEFT case the
effects  of heavier particles  are absorbed  into the  coefficients of
higher  dimension  operators. Since  the  momenta  are  allowed to  be
infinite, divergences  appear, and therefore the CEFT  need to undergo
both  regularization  and renormalization.  In  choosing the  specific
scheme  for  carrying out  the  above  procedures  particular care  is
needed. Whereas in principle one  could use any scheme, experience has
shown  that  the  most  natural   scheme  for  studying  the  CEFT  is
dimensional      regularization      with     minimal      subtraction
\cite{Weinberg:1979kz,Weinberg:1980wa,Hall:1981kf,Georgi:1993qn,
Leutwyler:1994iq,Manohar:1996cq,Pich:1998xt}. CEFT
are widely  used in Physics: for  example, when in the  context of QCD
one talks about 3, 4 or 5 active flavors, one is implicitly using this
latter type of  effective 
theories~\cite{Witten:1977kx,Weinberg:1980wa}. 
Moreover, 
most of  the analyses of
Grand Unification~\cite{Georgi:1974yf,Hall:1981kf} resort to 
CEFT-type of constructions: one has a full
theory at the GUT scale, then  an effective field theory below the GUT
scale  (SM or MSSM)  is built,  and then  yet another  effective field
theory  below the  Fermi  scale  (just QED+QCD).  In  these cases  the
complete theory is known, and the  CEFT language is used only in order
to simplify the calculations at  low energies and to control the large
logarithms   which   appear    when   there   are   widely   separated
scales. Nevertheless, CEFT's are  useful even when the complete theory
is not known,  or when the connection with  the complete theory cannot
be worked  out; this is  the case of  Chiral Perturbation theory  (\(
\chi PT \)) \cite{Callan:1969sn,Coleman:1969sm,
Weinberg:1968de,Gasser:1984yg} (for more recent reviews see
also \cite{Leutwyler:1994iq,Pich:1995bw,Ecker:1995gg}).

It is important  to maintain a sharp distinction  between the two types
of  EFT   mentioned  above,  i.e.  Wilsonian   or  continuum,  because
conceptually  they are quite  different. However,  perhaps due  to the
fact that the language is in part common to both types of theories, it
seems  that they  are often  used interchangeably  in  the literature,
especially  when  employing cutoffs  within  the  CEFT framework.   In
particular, since the  couplings \( \alpha _{i} \)  have dimensions \(
[\alpha _{i}]=M^{-n}  \), when  computing loops one  generally obtains
effects which  grow as \(  (\Lambda ^{n}\alpha _{i})^{m} \),  where \(
\Lambda \) is the formal CEFT  cutoff, and as such is void of physics.
As a  consequence, physical observables should be  made as independent
of these  cutoffs as possible  by introducing as many  counterterms as
needed   to    renormalize   the   answer.    Not   performing   these
renormalizations correctly, or identifying naively formal cutoffs with
the physical cutoffs  of the effective theory, can  lead to completely
non-sensical         results         (see         for         instance
\cite{Burgess:1992va,Burgess:1993gx}).  This  type of pitfalls  may be
avoided  by simply using  dimensional regularization, since  the latter
has  the  special property  of  not  mixing  operators with  different
dimensionalites.

The usual way to treat  theories with compactified extra dimensions is
to define them as a 4-dimensional  theory with a truncated tower of KK
modes at some  large but otherwise arbitrary \(  N_{s} \), a procedure
which effectively amounts to using a hard cutoff in the momenta of the
extra dimensions. Thus, physical  quantities calculated in this scheme
depend explicitly  on the  cutoff \( N_{s}  \), which  is subsequently
identified with  some physical cutoff. However,  as already commented,
\(  N_{s} \)  plays the  role  of a  formal cutoff,  and is  therefore
plagued with  all the  aforementioned ambiguities.  Identification of
this  formal cutoff  with a  universal  physical cutoff  can give  the
illusion of predictability, 
making us forget that we are
dealing  with  a non-renormalizable  theory  with  infinite number  of
parameters, which can be predictive only at low energies, where higher
dimension operators may be neglected.

In this paper we want to  analyze the question of the running of gauge
couplings  in   theories  with   compact  dimensions  from   the  CEFT
{}``canonical{}'' point of view. We  hasten to emphasize that even the 
CEFT
presents   conceptual   problems   in   theories   with   compactified
dimensions. Specifically, as mentioned above, in the CEFT approach the
(virtual) momenta are allowed to vary up to infinity; however, momenta
related to the compactified extra  dimensions turn out to be KK masses
in the  4-dimensional compactified theory,  where it is  supposed that
one  only  keeps particles  lighter  than  the  relevant scale.  Thus,
truncating the  KK series  amounts to cutting  off the momenta  of the
compactified  dimensions.   Therefore,  in  order  to  define  a  true
{}``non-cutoff{}'' CEFT scheme we are forced to keep all KK modes. Our
main motivation is to seriously explore this approach, and investigate
both its virtues and its limitations  for the problem at hand. We hope
that this  study will help  us identify more clearly  which quantities
can  and  which  cannot  be computed  in  effective  extra-dimensional
theories.

The paper is organized  as follows: In section \ref{sec:thresholds} we
discuss the  usual arguments  in favor of  power-law running  of gauge
couplings  and  show that  they  depend  crucially  on the  way  KK
thresholds  are  crossed.   In  particular  we show  that,  a  one-loop
$\beta$
function which  simply counts  the number of  modes, diverges  for more
than \(  5 \)  dimensions, if the  physical way of  passing thresholds
dictated by the vacuum polarization function (VPF) is imposed.

In section \ref{sec:model} we introduce  a theory with one fermion and
one  photon in  4+$\delta$ dimensions,  with the  extra  $\delta$ ones
compactified. This theory, which is  essentially QED in 4+\( \delta \)
dimensions, serves  as toy model  for studying the issue  of power
corrections and the running of the coupling in a definite framework.

In section  \ref{sec:effectiveKK} we study the  question of decoupling
KK modes in the aforementioned theory by analyzing the behavior of the
VPF of  the (zero-mode) photon. Since, as  commented above, decoupling
the KK modes  one by one is problematic, we study  the question of how
to decouple all  of them at once.  To accomplish  this we consider the
VPF of the photon with all KK modes included, and study how it reduces
at \(  Q^{2}\ll M_{c} \) to the  standard QED VPF with  only one light
mode.  Since the entire KK tower is kept untruncated, the theory is of
course non-renormalizable;  therefore, to compute  the VPF we  have to
regularize and renormalize it in the  spirit of the CEFT, in a similar
way that observables are  defined in \( \chi PT \).  As  in \( \chi PT
\),  it is  most  convenient to  use  dimensional regularization  with
minimal  subtraction, in  order to  maintain a  better control  on the
mixing  among  different  operators.  However,  at the  level  of  the
4-dimensional theory the non-renormalizability manifest itself through
the  appearance of  divergent sums  over  the infinite  KK modes,  and
dimensional regularization does no  seem to help in regularizing them.
The dimensional  regularization of the VPF  is eventually accomplished
by exploiting  the fact  that its UV
behavior coincides to that found when the 
\( \delta \)  extra dimensions have not been compactified\footnote{This
is in a way expected,
since for  very large \(Q^{2}\gg M_{c}=1/R_{c} \)
the compactification effects should be negligible. Note, however, that
this is not always the case; a known exception is provided by the orbifold
compactification~\cite{Contino:2001si}.}.
To explore this point
we first resort to  the standard unitarity relation (optical theorem),
which relates the imaginary part of the VPF to the total cross section
in the  presence of  the KK  modes; the latter is finite  because the
phase-space truncates the  series.  For $Q^2\gg M_c^2$ the uncompactified
result for the imaginary part of the VPF is rapidly reached, i.e. after 
passing a few thresholds. 
We then compute the real part of the one-loop VPF in
the non-compact theory in \(  4+\delta \) dimensions, where, of course
we can use directly dimensional regularization to regularize it (since
no  KK reduction has  taken place).   For later  use we  also present
results in which the same quantity is evaluated by using hard cutoffs.
Finally,  we show  that the  UV divergences  of the  one-loop  VPF are
indeed the  same in  both the (torus)-compactified  and uncompactified
theories.   Therefore,   in  order  to  regularize  the   VPF  in  the
compactified  theory  with  an  infinite  number of  KK  modes  it  is
sufficient  to split the  VPF into  two pieces,  an ``uncompactified''
piece,  corresponding  to the  case  where  the  extra dimensions  are
treated at the same footing as  the four usual ones, and a piece which
contains all compactification effects.  We show that this latter piece
is  UV  and  IR finite  and  proceed  to  evaluate  it, while  all  UV
divergences remain in the  former, which we evaluate using dimensional
regularization.

The     results     of    previous     sections     are    used     in
section~\ref{sec:matching}  to define  an effective  charge  \( \alpha
_{\mathrm{eff}}(Q) \)  which can be continuously  extrapolated from \(
Q^{2}\ll  M_{c} \) to  \( Q^{2}\gg  M_{c} \).   We use  this effective
charge to study the matching  of couplings in the low energy effective
theory (QED) to the couplings of the theory containing an infinite  of KK 
modes.  In
the context  of dimensional regularization we find  that this matching
contains only the standard logarithmic running from \( m_{Z} \) to the
compactification scale \( M_{c} \), with no power corrections.  On
the other hand, if hard cutoffs  are used to regularize the VPF in the
non-compact  space,  one  does  find power corrections,
which may  be interpreted  as  an  additional
matching between  the effective \( D=4+\delta \) dimensional field  theory and
some  more complete  theory. We discuss the possibility of estimating this
matching in the EFT without knowing the details of the full theory. This
point is studied in a simple extension 
of our original toy-model, by endowing 
the theory considered (QED in  4+\( \delta \) compact dimensions) with
an  additional  fermion with  mass  \(  M_{f}\gg  M_{c} \),  which  is
eventually integrated out.

\section{Crossing Thresholds\label{sec:thresholds}}

The simplest argument (apart from the purely dimensional ones)
in favor of power-law running in theories with extra dimensions is based on
the fact that in \( \msb  \)-like schemes the $\beta$ function is
proportional to the number of active modes.
Theories with \( \delta  \) extra 
compact dimensions contain, in general, 
a tower of KK modes. In particular, if we embed QED in extra dimensions
we find that electrons (also photons) 
have a tower of KK modes 
with masses \( M^{2}_{n}=\left( n^{2}_{1}+n^{2}_{2}+
\cdots +n^{2}_{\delta }\right)  \)
\( M^{2}_{c} \) with \( n_{i} \) integer values and \( M_{c}=1/R_{c} \) the
compactification scale. The exact multiplicity of the spectrum depends on the
details of the compactification procedure (torus, orbifold, etc). As soon as
we cross the 
compactification scale, the KK modes begin to contribute,
and therefore one expects that the $\beta$ function 
of this theory will start to
receive additional contributions from them. In a general renormalization scheme
satisfying decoupling one can naively write 
\begin{equation}
\label{eq:beta-naive}
\beta =\sum _{n}\beta _{0}f\left( \frac{\mu }{M_{n}}\right)~, 
\end{equation}
 where \( \mu  \) is the renormalization scale, 
\( \beta _{0} \) is the contribution
of a single mode, and \( f(\mu /M) \) is a general step-function that decouples
the modes as \( \mu  \) crosses the different thresholds, namely \( f(\mu /M)\rightarrow 0\, \, \, \, \, \mu \ll M \)
and \( f(\mu /M)\rightarrow 1\, \, \, \, \, \mu \gg M \). For instance in \( \msb  \) schemes \( f(\mu /M)\equiv \theta (\mu /M-1) \) 
where \( \theta (x) \) is
the step-function. Then one finds \( \left( \Omega _{\delta }=2\pi ^{\delta /2}/\Gamma (\delta /2)\right)  \)
\begin{equation}
\label{eq:beta-naive-dr}
\beta =\sum _{n<\mu /M_{c}}\beta _{0}\approx \beta _{0}\int d\Omega _{\delta
}n^{\delta -1}dn=\beta _{0}\frac{\Omega _{\delta }}{\delta }\left( \frac{\mu ^{2}}{M^{2}_{c}}\right) ^{\delta /2}~.
\end{equation}
 This argument, simple and compelling as it may seem, cannot be trusted completely
because in \( \msb  \) schemes the decoupling is put in by hand.
Therefore, other types of schemes, in which decoupling seems natural, have been
studied in the literature. For instance, in ref.~\cite{Dienes:1998vg} the
VPF of the photon at \( Q^{2}=0 \) was calculated in the presence of the infinite
tower of KK modes by using a hard cutoff in proper time, and the result was
used to compute the $\beta$ function; in that case the modes decouple smoothly.
In addition, after adjusting the cutoff by hand one can reproduce the aforementioned
result obtained in \( \msb  \). One can easily see that this procedure is equivalent
to the use of the function \( f(\Lambda /M)\equiv e^{-\frac{M^{2}_{n}}{\Lambda ^{2}}} \)
to decouple the KK modes \begin{equation}
\label{eq:beta-naive-kk}
\beta =\sum _{n}\beta _{0}e^{-\frac{M^{2}_{n}}{\Lambda ^{2}}}\approx \beta _{0}\left( \pi \frac{\Lambda ^{2}}{M^{2}_{c}}\right) ^{\delta /2}\, .
\end{equation}
 If one chooses by hand 
 \( \mu ^{\delta }=\Gamma (1+\delta /2)\Lambda ^{\delta } \),
the sum in Eq.~(\ref{eq:beta-naive-kk}) agrees exactly with the sum obtained
if one uses a sharp step-function. Even though this particular way of decoupling
KK modes appears naturally in some string 
scenarios~\cite{Quiros:2001yi,Hamidi:1987vh,Dixon:1987qv,Antoniadis:1994jp},
it hardly appears compelling from the field theory point of view; this procedure
is not any better conceptually than the sharp step-function decoupling of modes:
one obtains a smooth $\beta$ function because one uses a smooth function to decouple
the KK modes.

These two ways of decoupling KK modes, due to the very sharp step-like behavior
they impose, lead to a finite result in (\ref{eq:beta-naive}) for any number
of extra dimensions. One is tempted to ask, however, what would happen if one
were to use a more physical way of passing thresholds. In fact, heavy particles
decouple naturally and smoothly in the VPF, because they cannot be
produced physically. Specifically, in QED in 4-dimensions at the one-loop level,
the imaginary part, \( \Im m\Pi (q^{2}) \), of the VPF \( \Pi (q^{2}) \) is
directly related, via the optical theorem, to the tree level cross sections
\( \sigma  \) for the physical processes \( e^{+}e^{-}\rightarrow f^{+}f^{-} \),
by 
\begin{equation}
\label{sigmaff1}
\Im m\Pi (s)=\frac{s}{e^{2}}\, \sigma (e^{+}e^{-}\rightarrow f^{+}f^{-})~.
\end{equation}
Given a particular contribution to the spectral function \( \Im m\Pi (s) \),
the corresponding contribution to the renormalized vacuum polarization function
\( \Pi _{R}(q^{2}) \) can be reconstructed via a once--subtracted dispersion
relation. For example, for the one--loop contribution of the fermion \( f \),
choosing the on--shell renormalization scheme, one finds (if \( q \) is the
physical momentum transfer with \( q^{2}<0 \), as usual we define \( Q^{2}\equiv -q^{2} \)):
\[
\Pi _{R}(Q)=Q^{2}\, \int _{4m_{f}^{2}}^{\infty }ds\frac{1}{s(s+Q^{2})}\frac{1}{\pi }\, \Im m\Pi (s)\]

\begin{equation}
\label{PiRffQED}
=\frac{\alpha }{\pi }\times \left\{ \begin{array}{lr}
\displaystyle {\frac{1}{15}\frac{Q^{2}}{m_{f}^{2}}+\mathcal{O}\left( \frac{Q^{4}}{m_{f}^{4}}\right) } & \, \, \, \, Q^{2}/m_{f}^{2}\rightarrow 0\\
{\frac{1}{3}\, \ln \left( \frac{Q^{2}}{m^{2}_{f}}\right) -\frac{5}{9}+\mathcal{O}\left( \frac{m_{f}^{2}}{Q^{2}}\right) } & \, \, \, \, Q^{2}/m_{f}^{2}\rightarrow \infty \, ,
\end{array}\right. 
\end{equation}
where \( \alpha \equiv e^{2}/(4\pi ) \). The above properties can be extended
to the QCD effective charge \cite{Binosi:2002vk}, with the
appropriate modifications to take into account the non-abelian nature of the
theory, and provide a physical way for computing the matching equations between
couplings in QCD at quark mass thresholds. One computes the VPF of QCD with
\( n_{f} \) flavors and that of QCD with \( n_{f}-1 \) flavors, and requires
that the effective charge is the same for \( Q^{2}\ll m_{f} \) in the two theories.
This procedure gives the correct relation between the couplings in the two theories
~\cite{Bernreuther:1982sg,Rodrigo:1993hc,Rodrigo:1998zd}. However, one can
easily see that this cannot work for more than one extra dimension. To see that,
let us consider the decoupling function \( f(\mu /M) \) provided by the
one-loop VPF, which, as explained, captures correctly the physical thresholds.
The corresponding \( f(\mu /M) \) may be obtained by differentiating \( \Pi
_{R}(Q) \)
once with respect to \( Q^{2} \); it is known \cite{Brodsky:1998mf} that the
answer can be well-approximated by a simpler function of the form \( f(\mu /M)=\mu ^{2}/(\mu ^{2}+5M^{2}) \).
We see immediately that if we insert this last function in
Eq.~(\ref{eq:beta-naive})
and perform the sum over all KK modes the result is convergent only for one
extra dimension (with a coefficient which is different from the one obtained
with the renormalization schemes mentioned earlier), while it becomes highly divergent for
several extra dimensions. We conclude therefore that the physical way of decoupling
thresholds provided by the VPF seems to lead to a divergent $\beta$ function in
more than one extra dimension. As we will see, this is due to the fact that,
in order to define properly 
the one-loop VPF for \( \delta >1 \), more than one subtraction is needed.

\section{A Toy Model\label{sec:model}}

To be definite we will consider a theory 
with one fermion and one photon in
\( 4+\delta  \) dimensions, 
in which the \( \delta  \) extra dimensions are
compactified on a torus of 
equal radii \( R_{c}\equiv 1/M_{c} \). 
The Lagrangian is given by
\begin{equation}
\label{eq:DeffLagrangian}
{\mathcal{L}}_{\delta }=-\frac{1}{4}F^{MN}F_{MN}+i\bar{\psi }\gamma
^{M}D_{M}\psi +\mathcal{L}_{\mathrm{ct}}~,
\end{equation}
where \( M=0,\cdots ,3,\cdots ,3+\delta  \). We will also use greek letters
to denote four-dimensional indices \( \mu =0,\cdots \, 3 \). \( D_{M}=\partial _{M}-ie_{D}A_{M} \)
is the covariant derivative with \( e_{D} \) the coupling in \( 4+\delta  \)
dimensions which has dimension \( [e_{D}]=1/M^{\delta /2} \). 
After compactification, the dimensionless
gauge coupling in four-dimensions, 
\( e_{4} \), 
and the dimensionfull \( 4+\delta  \)
coupling are related by the compactification scale \footnote{%
Note that the factors \( 2\pi  \) depend on the exact way the extra dimensions
are compactified (on a circle, orbifold, etc). }
\begin{equation} 
e_{4}=e_{D}\left( \frac{M_{c}}{2\pi }\right) ^{\delta /2}  \,. 
\label{relcoup}
\end{equation}
Evidently 
\( e_{D} \) is determined from 
the four-dimensional gauge coupling and the compactification
scale, but in the uncompactified space we can regard it  
as a free parameter
(as \( f_{\pi } \) in \( \chi PT \)). 
Finally \( \mathcal{L}_{\mathrm{ct}} \)
represents possible gauge invariant operators 
with dimension \( 2+D \) or higher,
which are in general needed for renormalizing the theory; they can be computed
only if a more complete theory, from which our effective theory originates,
is given. For instance, by computing the VPF we will see that 
a $\mathcal{L}_{\mathrm{ct}}$ of the form 
\begin{equation}
\label{eq:lct}
\mathcal{L}_{\mathrm{ct}}=\frac{c_{1}}{M^{2}_{s}}D_{M}F^{MK}D^{N}F_{NK}+\cdots
\end{equation}
is needed to make it finite. 

The spectrum after compactification contains a photon (the zero mode of the
four-dimensional components of the gauge boson), the \( \delta  \) extra 
components
of the gauge boson remain in the spectrum as \( \delta  \) massless real scalars,
a tower of massive vector bosons with masses \( M^{2}_{n}=\left( n^{2}_{1}+n^{2}_{2}+\cdots +n^{2}_{\delta }\right)  \)
\( M^{2}_{c} \), \( n_{i}\in \mathbb {Z},\, n_{i}\neq 0 \) , \( 2^{[\delta /2]} \)
massless Dirac fermions (here the symbol \( [x] \) represents the closest integer
to \( x \) smaller or equal than \( x \)), and 
a tower of massive Dirac fermions
with masses given also by the above mass formula. Note that this theory does
not lead to normal QED at low energies, first because the \( \delta  \) extra
components of the gauge boson remain in the spectrum, and second because in
\( 4+\delta  \) dimensions the fermions have \( 4\cdot 2^{[\delta /2]} \)
components, which remain as zero modes, leading at low energy to a theory with
\( 2^{[\delta /2]} \) Dirac fermions. In the \( D=4+\delta  \) theory these
will arise from the trace of the identity of the $\gamma$ matrices, which just
counts the number of components of the spinors. To obtain QED as a low energy
one should project out the correct degrees of 
freedom by using some more appropriate
compactification (for instance, orbifold compactifications can remove the extra
components of the photon from the low energy spectrum, and leave just one Dirac
fermion). However this is not important for our discussion of the VPF, we just
have to remember to drop the additional factors \( 2^{[\delta /2]} \) to make
contact with usual QED with only one fermion. Theories of this type, with all
particles living in extra dimensions are called theories with {}``universal
extra dimensions{}''~\cite{Appelquist:2000nn} and have the characteristic
that all the effects of the KK modes below the compactification scale cancel
at tree level due to the conservation of the KK numbers. 
In particular, and contrary to what happens in 
theories where gauge and scalar fields 
live in the bulk and fermions in the 
brane~\cite{Pomarol:1998sd,Delgado:1999sv}, 
no divergences associated to summations over KK towers 
appear at tree level.  
Finally, the couplings
of the electron KK modes to the standard zero-mode photon are universal and
dictated by gauge invariance. The couplings among the KK modes can be found
elsewhere \cite{Papavassiliou:2001be,Muck:2001yv}; they will not be important
for our discussion of the VPF that we present here.

\section{The vacuum polarization in the presence of \protect\( KK\protect \)
modes\label{sec:effectiveKK}}

In this section we will study in detail the behavior of the 
one-loop VPF in the 
theory defined above for general values of the 
number $\delta$ of extra dimensions 
The main problems we want to 
address are: i) the general 
divergence structure of the VPF, ii) demonstrate that it is 
possible to regulate the UV divergences using dimensional regularization,
iii) the appearance of non-logarithmic 
(power) corrections, and, iv) their comparison
to the analogous terms obtained when resorting to a hard-cutoff 
regularization.

\subsection{The imaginary part of the vacuum polarization}

One can try to compute directly the VPF of the zero-mode photon in a theory
with infinite KK fermionic modes. However, one immediately sees that, 
in addition
to the logarithmic divergences that one finds in QED, new divergences
are encountered when summing over the infinite number of KK modes. One can understand
the physical origin of these divergences more clearly by resorting to the unitarity
relation (here \( s \) denotes the center-of-mass energy 
available for the production process):
\begin{eqnarray}
\Im m\Pi ^{(\delta )}(s) & = & \frac{s}{e_{4}^{2}}\, \sum _{n}\sigma (e^{+}e^{-}\rightarrow f_{n}^{+}f_{n}^{-})\nonumber \label{eq:first} \\
 & = & \frac{\alpha _{4}}{3}\sum _{n<n_{\mathrm{th}}}\left( 1+\frac{2M_{n}^{2}}{s}\right) \sqrt{1-4M_{n}^{2}/s}\, \, ,
\end{eqnarray}
 where \( n<n_{\mathrm{th}} \) represents the sum 
over all the electron KK
modes such that 
\( 4\left( n^{2}_{1}+n^{2}_{2}+\cdots n^{2}_{\delta }\right) M^{2}_{c}<s \), 
and \( \alpha _{4}=e^{2}_{4}/(4\pi ) \). 
This sum can be evaluated approximately
for \( s\gg M^{2}_{c} \) by replacing it by an integral; then we obtain

\begin{equation}
\label{QWER}
\Im m\Pi ^{(\delta )}(s)\approx \frac{\alpha _{4}}{2^{3+\delta }}\frac{(\delta +2)\, \pi ^{(\delta +1)/2}}{\Gamma \left( (\delta +5)/2\right) }\left( \frac{s}{M^{2}_{c}}\right) ^{\delta /2} \,.
\end{equation}

\begin{figure}
{\par\centering \includegraphics{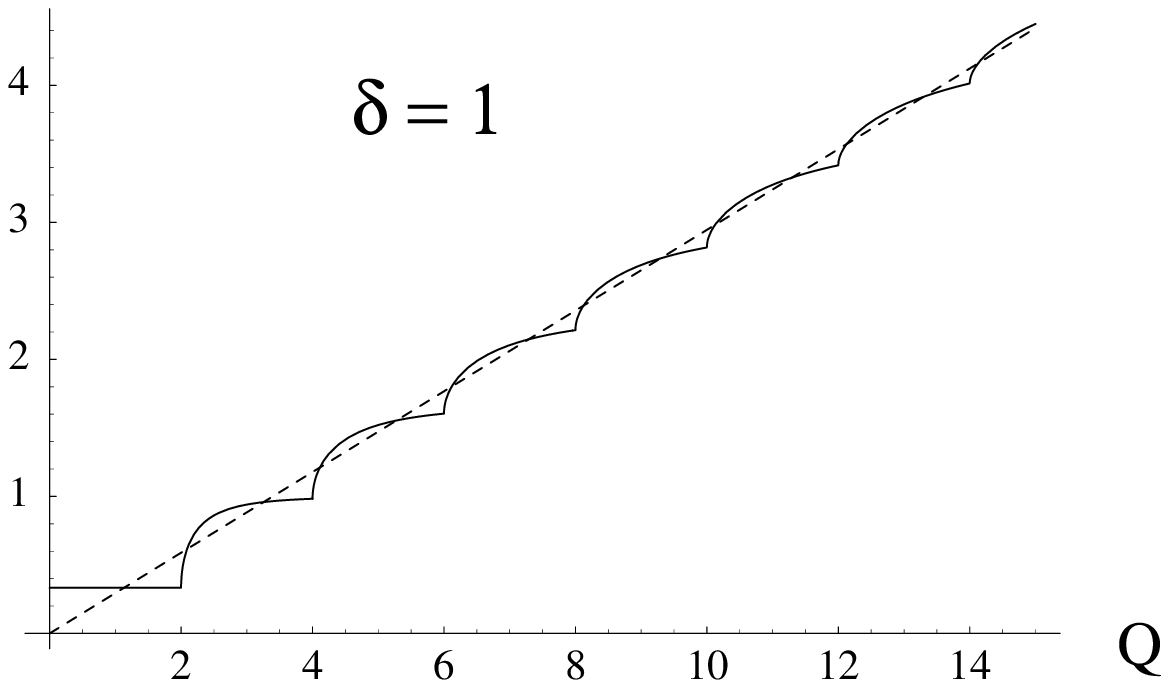} \par}

{\par\centering \includegraphics{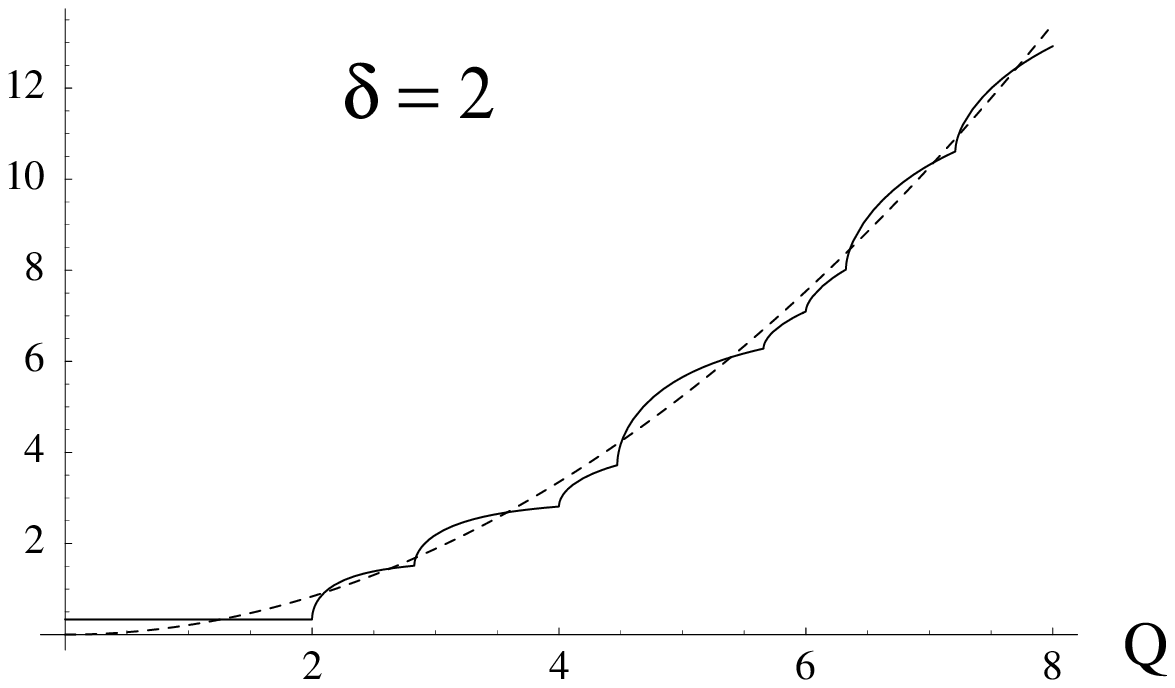} \par}

{\par\centering \includegraphics{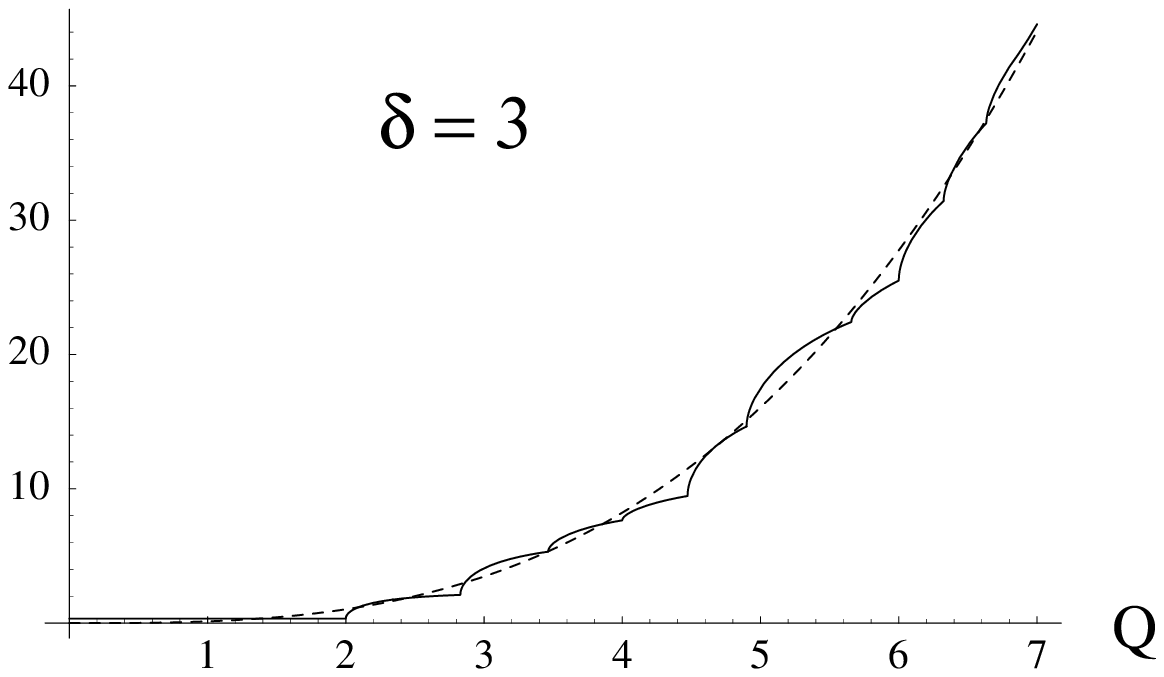} \par}

\caption{\label{fig:imaginary}\protect\( \Im m\Pi ^{(\delta )}(Q)\protect \) as compared
with the asymptotic value (\protect\( \delta =1,2,3\protect \)). $Q$ is
given in units of $M_c$.}
\end{figure}

It turns out that this last result 
captures the behavior of the 
same quantity when
the extra dimensions are not compact; this is so  
because, at high energies, the effects of the compactification
can be neglected. 
In fact, this result may be deduced on simple dimensional grounds: as
commented, the gauge coupling in \( 4+\delta  \) dimensions has dimension \( 1/M^{\delta /2} \);
therefore one expects that \( \Im m\Pi ^{(\delta )}(s) \) will grow with \( s \)
as \( \left( s/M^{2}\right) ^{\delta /2} \), which is what we obtained from
the explicit calculation. To see how rapidly one reaches this regime we can
plot the exact result of \( \Im m\Pi ^{(\delta )}(s) \) together with the asymptotic
value.
As we can 
see in Fig.\ref{fig:imaginary}, the asymptotic limit is reached very
fast, especially for higher dimensions. 
For practical purposes one can reliably use the asymptotic
value soon after passing the first threshold, \( Q>2M_{c} \), 
incurring errors which are below 10\%.

Now we can try to obtain the real part by using a dispersion relation as the
one used in 4-dimensional QED, i.e. Eq.(\ref{PiRffQED}). However,
one immediately
sees that it will need a number of subtractions which depends 
on the value of \( \delta\).
Thus, for just one extra dimension, as in 4-dimensional QED, one subtraction
is enough, for \( \delta =2 \) and \( \delta =3 \) two subtractions are needed,
and so on. This just manifests the non-renormalizability of the theory, and
in the effective field theory language, the need for higher dimension operators
acting as counterterms. 
Even though this ``absorptive'' approach is perfectly acceptable,  
it would be preferable to have a way of computing
the real part directly at the Lagrangian level 
(by computing loops, for instance).
As commented in the introduction, to accomplish this  
we will use dimensional regularization.

\subsection{The vacuum polarization in uncompactified \protect\( 4+\delta \protect \)
dimensions}

When using dimensional regularization to compute the VPF in uncompactified
space, to be denoted $\Pi_{\mathrm{uc}}$, simple dimensional arguments suggest
that one should typically obtain contributions of the form
\[
\Pi _{\mathrm{uc}}(Q)\propto e^{2}_{4}\, \left( 2\pi \frac{Q}{M_{c}}\right)
^{\delta }\, ~,
\]
since the two vertices in the loop provide a factor \( e^{2}_{D} \), whose
dimensions must be compensated by the only 
available scale in the problem, 
namely
\( Q^{2} \). In the above formula we have 
used the relation of Eq.(\ref{relcoup})
in order to trade off \( e_{D} \)
for  \( e_{4} \). The omitted 
coefficient in front will be generally divergent, 
and will be regularized by letting
\( \delta \rightarrow \delta -\epsilon  \).

Let us compute the VPF \( \Pi ^{MN}_{\mathrm{uc}}(q) \) 
in uncompactified space, assuming that, if necessary, 
the dimensions will be continued to complex values.
We have that 
\begin{equation}
\Pi ^{MN}_{\mathrm{uc}}(q)=ie^{2}_{D}\int \frac{d^{4+\delta }k}{(2\pi
)^{4+\delta }}\mathrm{Tr}\left\{ \gamma ^{M}\frac{1}{\sla k}\gamma
^{N}\frac{1}{\sla k+\sla q}\right\}~,
\end{equation}
which, by gauge-invariance assumes the standard form 
\[
\Pi ^{MN}_{\mathrm{uc}}(q)=\left( q^{2}g^{MN}-q^{M}q^{N}\right) \Pi
_{\mathrm{uc}}(q)~.
\]
If we now were to use that, in \( D \)-dimensions,  
Tr\( [\gamma ^{M}\gamma ^{N}]=2^{[D/2]}g^{MN} \), 
we would find that the low energy limit has an extra \( 2^{[\delta /2]} \)
factor, which, as commented, is an artifact of the torus compactification: there
are \( 2^{[\delta /2]} \) too many fermions in the theory. Therefore we simply
drop this factor by hand. Moreover, 
we use Eq.(\ref{relcoup})
and employ the proper-time parametrization in intermediate steps,
thus arriving at:
\begin{eqnarray}
\Pi _{\mathrm{uc}}(Q) & = & \frac{e^{2}_{4}}{2\pi ^{2}}\left( \frac{\pi }{M_{c}^{2}}\right) ^{\delta /2}\int _{0}^{1}dxx(1-x)\int ^{\infty }_{0}\frac{d\tau }{\tau ^{1+\frac{\delta }{2}}}\exp \left\{ -\tau \, x(1-x)Q^{2}\right\} \nonumber \\
 & = & \frac{e^{2}_{4}}{2\pi ^{2}}\frac{\pi ^{\delta /2}\, \Gamma
^{2}(2+\frac{\delta }{2})}{\Gamma (4+\delta )}\, \Gamma \left( -\frac{\delta }{2}\right) \, \left( \frac{Q^{2}}{M_{c}^{2}}\right) ^{\delta /2}~.
\label{EFC1}
\end{eqnarray}

A simple check of this result may be obtained by computing its imaginary part.
To that end we let \( Q^{2}\rightarrow -q^{2}-i\epsilon  \) with \( q^{2}>0 \).
Then \[
\Im m\left\{ -q^{2}-i\epsilon \right\} ^{\delta /2}=-\left( q^{2}\right) ^{\delta /2}\sin \frac{\delta \pi }{2}\, .\]
 Now we can use that \( \Gamma (-\delta /2)\Gamma (1+\delta /2)=-\pi /\sin (\delta \pi /2) \)
to write

\[
\Im m\left\{ \Pi _{\mathrm{uc}}(q)\right\} =\alpha _{4}\frac{2\pi ^{\delta /2}\, \Gamma ^{2}(2+\frac{\delta }{2})}{\Gamma (4+\delta )\Gamma (1+\delta /2)}\left( \frac{q^{2}}{M_{c}^{2}}\right) ^{\delta /2}=\frac{\alpha _{4}}{2^{3+\delta }}\frac{(\delta +2)\, \pi ^{(\delta +1)/2}}{\Gamma \left( (\delta +5)/2\right) }\left( \frac{q^{2}}{M^{2}_{c}}\right) ^{\delta /2}\, \]
which agrees with our previous result of Eq.(\ref{QWER}).

For odd values of \( \delta  \), the one-loop \( \Pi _{\mathrm{uc}}(Q) \) computed
above is finite, since the \( \Gamma (-\frac{\delta }{2}) \) can be calculated
by analytic continuation. This result is in a way expected, since in odd number
of dimensions, by Lorentz invariance, there are no appropriate gauge invariant
operators able to absorb any possible infinities generated in the one-loop VPF;
this would require operators which give contributions that go like \( Q^{\delta } \).
Notice, however, that at higher orders \( \Pi _{\mathrm{uc}}(Q) \) will eventually
become divergent. For instance, in five dimensions at two loops, the VPF should
go as \( Q^{2} \), since there are four elementary vertices. The divergences
generated by these contributions could be absorbed in an operator such as the
one considered in the previous section, namely \( D_{M}F^{MK}D^{N}F_{NK} \).
On the other hand, when \( \delta  \) is even, \( \Gamma (-\frac{\delta }{2}) \)
has a pole, and subtractions are needed already at one loop. To compute the
divergent and finite parts in a well-defined way we will use dimensional regularization,
i.e. we will assume that \( \delta \rightarrow \delta -\epsilon  \) . 
Notice however that, unlike in 4-dimensions,  
we not need to introduce an additional scale at this point, 
i.e. the equivalent of the 't Hooft mass scale $\mu$: 
\( M_{c} \) plays
the role of \( \mu  \), and can be used to keep \( e_{4} \) dimensionless. 
After expanding
in \( \epsilon  \) we find a simple pole accompanied by the usual logarithm
\begin{equation}
\label{eq:uc-divergent}
\Pi _{\mathrm{uc}}(Q)\propto \left( \frac{Q^{2}}{M_{c}^{2}}\right) ^{\delta /2}\left\{ -\frac{2}{\epsilon }+\ln (Q^{2}/M_{c}^{2})+\cdots \right\} \, .
\end{equation}
 Here the ellipses represent a finite constant. Now, to renormalize this result
we must introduce higher dimension operators (for instance, if \( \delta =2 \)
the operator \( D_{M}F^{MK}D^{N}F_{NK} \) will do the job) which could absorb
the divergent piece. The downside of this, however, is that we also have to
introduce
an arbitrary counterterm, \( \kappa  \), corresponding to the contribution
of the higher dimension operator; thus 
we obtain a finite quantity proportional
to \( \log (Q^{2}/M^{2}_{c})+\kappa  \). Note that, since \( \kappa  \) is
arbitrary, we can always introduce back a renormalization scale and write \( \log (Q^{2}/M^{2}_{c})+\kappa =\log (Q^{2}/\mu ^{2})+\kappa (\mu ) \)
with \( \kappa (\mu )=\kappa +\log (\mu ^{2}/M^{2}_{c}) \). It is also important
to remark that, in the case of odd number of dimensions, although at one loop
we do not need any counterterm to make the VPF finite, higher dimensional operators
could still be present and affect its value.

In the case of uncompactified space, 
it is interesting to compare the above result with that 
obtained  
by regularizing the integral using a hard cutoff. To study this it
is enough to carry out 
the integral of Eq.~(\ref{EFC1}), with a cutoff in 
\( \tau_{0}=1/\Lambda ^{2} \): 
\begin{equation}
\label{eq:PiCutoff}
\Pi_{\mathrm{uc}}(Q)=\frac{e_{4}^{2}}{2\pi ^{2}}\left( \frac{\pi }{M_{c}^{2}}\right) ^{\delta /2}\int _{0}^{1}dx\, x(1-x)\int _{\tau _{0}}^{\infty }\frac{d\tau }{\tau ^{1+\delta /2}}\exp \left\{ -\tau \, x(1-x)Q^{2}\right\} .
\end{equation}
 Then, for \( \delta =1,2,3 \) we obtain
\begin{equation}
\Pi ^{(1)}_{\mathrm{uc}}(Q)=\frac{e_{4}^{2}}{2\pi ^{2}}\left( -\frac{3\pi
^{2}Q}{64M_{c}}+\frac{\sqrt{\pi }Q^{2}}{15M_{c}\Lambda }+\frac{\sqrt{\pi
}\Lambda }{3M_{c}}\right) \, ~,
\label{d1}
\end{equation}
\begin{equation}
\Pi ^{(2)}_{\mathrm{uc}}(Q)=\frac{e_{4}^{2}}{2\pi ^{2}}\left( \frac{\pi
\Lambda ^{2}}{6M_{c}^{2}}+\frac{\pi Q^{2}}{30M_{c}^{2}}\left( \log
(Q^{2}/\Lambda ^{2})+\gamma -\frac{77}{30}\right) \right) \, ~,
\label{d2}
\end{equation}
\begin{equation}
\Pi ^{(3)}_{\mathrm{uc}}(Q)=\frac{e_{4}^{2}}{2\pi ^{2}}\left( \frac{5\pi
^{3}Q^{3}}{768M_{c}^{3}}-\frac{\pi ^{3/2}Q^{2}\Lambda
}{15M_{c}^{3}}+\frac{\pi ^{3/2}\Lambda ^{3}}{9M_{c}^{3}}\right) \, ~.
\label{d3}
\end{equation}

As we can see, 
the pieces which are independent of the cutoff are exactly 
the same ones we obtained
using dimensional regularization. But, in addition, we obtain a series of
contributions which depend explicitly on the cutoff. 
For instance we find 
corrections to the gauge coupling which behave 
as \( \Lambda ^{\delta } \),
and just redefine the gauge coupling we started with \cite{Taylor:1988vt}. 
In
the case of 5 dimensions we also generate a term linear in \( Q^{2} \); however
it is suppressed by 1/\( \Lambda  \), and therefore it approaches zero for
large \( \Lambda  \). In the case of 6 dimensions we obtain the same logarithmic
behavior we found with dimensional regularization, and the result can be cast
in identical form, if the cutoff is absorbed in the appropriate counterterm.
For 7 dimensions we also find divergent contributions which go as \( Q^{2} \).
This means that, when using cutoffs, 
higher dimension operators in the derivative expansion (e.g. operators
giving contributions as \( Q^{2} \) or higher)
are necessary to renormalize the theory and must be included. In the case of
dimensional regularization this type of operators is not strictly needed at
one loop; however, nothing forbids them in the Lagrangian, 
and they could appear
as ``finite counterterms''. If one were to identify the \( \Lambda  \)
in the above expressions with a physical cutoff, 
one might get the impression
that, contrary to the dimensional regularization 
approach where arbitrary counterterms
are needed,  one could now obtain all types of contributions
with only one additional parameter, namely \( \Lambda  \).
This is however not true: the regulator
function is arbitrary, we simply have chosen one among an infinity of possibilities.
By changing the regulator function we can change the coefficients of the different
contributions at will, except for those few contributions 
which are independent
of \( \Lambda  \). These latter are 
precisely the ones we have obtained
by using dimensional regularization. Thus, even when using cutoffs one has to
add counterterms from higher dimension operators, absorb the cutoff, and express
the result in terms of a series of unknown coefficients. The lesson is that
with dimensional regularization we obtain all calculable pieces, while the non-calculable
pieces are related to higher dimensional terms in the Lagrangian.

What we will demonstrate next is that the one-loop VPF in the compactified theory
on a torus can be renormalized exactly as the VPF in the uncompactified theory;
this will allow us to compute it for any number of dimensions, and examine its
behavior for large and for small values of the \( Q^{2} \).

\subsection{The vacuum polarization in \protect\( \delta \protect \) compact dimensions}

From the four-dimensional point of view the vacuum polarization tensor in the
compactified theory is
\[
\Pi^{\mu \nu }(q^2)=\sum _{n}ie_{4}^{2}\int \frac{d^{4}k}{(2\pi )^{4}}\mathrm{Tr}\left\{ \gamma ^{\mu }\frac{1}{\sla k-m_{n}}\gamma ^{\nu }\frac{1}{\sla k+\sla q-m_{n}}\right\} 
\]
with 
\( m^{2}_{n}=\left( n^{2}_{1}+n^{2}_{2}+\cdots +n^{2}_{\delta }\right) M^{2}_{c} \);
for simplicity we have assumed a common compactification radius \( R=1/M_{c} \)
for all the extra dimensions. The sum over \( n \) denotes collectively the
sum over all the modes \( n_{i}=-\infty ,\cdots ,+\infty  \). 
The contribution of each mode
to this quantity seems quadratically divergent, like in ordinary QED; however,
we know that gauge invariance converts it to only logarithmically divergent.
But, in addition, the sum over all the modes makes the above expressions highly
divergent. Instead of attempting to compute it directly,
we will add and subtract
the contribution of the vacuum polarization function of the uncompactified theory
in \( 4+\delta  \) dimensions:
\begin{equation}
\label{eq:VPFSplitting}
\Pi ^{\mu \nu }(q)=
\left[\Pi ^{\mu \nu }(q)-\Pi _{\mathrm{uc}}^{\mu\nu }(q)\right]+
\Pi_{\mathrm{uc}}^{\mu\nu }(q)=
\Pi _{\mathrm{fin}}^{\mu \nu }(q)+\Pi _{\mathrm{uc}}^{\mu\nu }(q)~.
\end{equation}
Here we have taken already into account the relation between the coupling in
\( 4+\delta  \) dimensions and the four-dimensional coupling and have restricted
the external Lorentz indices to the 4-dimensional ones. Depending on the value
of \( \delta  \) the vacuum polarization can be highly divergent 
(naively as \( \Lambda ^{\delta +2} \), 
and after taking into account gauge invariance
as \( \Lambda ^{\delta } \)). However, we can use dimensional regularization
(or any other regularization scheme) to make it finite. The important point
is that the quantity \( \Pi _{\mathrm{fin}}^{\mu \nu }(Q) \) is UV and IR finite
and can unambiguously computed.

Instead of doing the two calculations from scratch, we will do the following:

i) We will first compute the compactified expression
by using Schwinger's proper
time,~$\tau$, to regularize the UV divergences.

ii) We will show that the UV behavior of the compactified theory, \( \tau \rightarrow 0, \)
is just the behavior of the uncompactified theory.

iii) Therefore, to compute \( \Pi _{\mathrm{fin}}^{\mu \nu }(Q) \) it is sufficient
to compute \( \Pi ^{\mu \nu }(Q) \) and then subtract its most divergent contribution
when \( \tau \rightarrow 0 \). 
We will see that it is sufficient to make it finite.

After a few manipulations \( \Pi ^{\mu \nu }(Q) \) can be written as 
\[
\Pi ^{\mu \nu }(q)=\left( q^{2}g^{\mu \nu }-q^{\mu }q^{\nu }\right) \Pi (q) ~,
\]
 where \cite{Dienes:1998vg}
\[
\Pi (Q)=\frac{e_{4}^{2}}{2\pi ^{2}}\sum _{n}\int ^{1}_{0}dxx(1-x)\int
^{\infty }_{0}\frac{d\tau }{\tau }\exp \left\{ -\tau \left( x(1-x)Q^{2}+m^{2}_{n}\right) \right\} ~.
\]
 \( \Pi (Q) \) can be written in terms of the function
\[
\bar{\theta }_{3}(\tau )\equiv \sum ^{+\infty }_{n=-\infty }e^{-n^{2}\tau }=\sqrt{\frac{\pi }{\tau }}\bar{\theta }_{3}\left( \frac{\pi ^{2}}{\tau }\right) 
\]
 as
\[
\Pi (Q)=\frac{e_{4}^{2}}{2\pi ^{2}}\int ^{1}_{0}dxx(1-x)\int ^{\infty
}_{0}\frac{d\tau }{\tau }\exp \left\{ -\tau
x(1-x)\frac{Q^{2}}{M^{2}_{c}}\right\} \bar{\theta }^{\delta }_{3}(\tau )~,
\]
where we have rescaled \( \tau  \) in order to remove \( M_{c} \) from the
\( \bar{\theta }_{3}(\tau ) \) function. This last expression for \( \Pi (Q) \)
is highly divergent in the UV (\( \tau \rightarrow 0) \), because in that limit
the \( \bar{\theta }_{3}(\tau ) \) function goes as \( \sqrt{\pi /\tau } \).
Then, if we define, as in Eq.~(\ref{eq:VPFSplitting}),
$\Pi_{\mathrm fin}=\Pi-\Pi_{\mathrm uc}$, we have
\[
\Pi _{\mathrm{fin}}(Q)=\frac{e_{4}^{2}}{2\pi ^{2}}\int ^{1}_{0}dxx(1-x)\int
^{\infty }_{0}\frac{d\tau }{\tau }\exp \left\{ -\tau
x(1-x)\frac{Q^{2}}{M^{2}_{c}}\right\} \left( \bar{\theta }^{\delta }_{3}(\tau )-\left( \frac{\pi }{\tau }\right) ^{\delta /2}\right)~,
\]
which is completely finite for any number of dimensions. In fact,
the last term provides a factor 
\[
F_{\delta }(\tau )\equiv \bar{\theta }^{\delta }_{3}(\tau )-\left( \frac{\pi
}{\tau }\right) ^{\delta /2}\stackrel{\tau \rightarrow 0}{\longrightarrow
}2\delta \left( \frac{\pi }{\tau }\right) ^{\delta /2}\exp \left\{ -\frac{\pi ^{2}}{\tau }\right\}~,
\]
that makes the integral convergent in the UV, while for large \( \tau  \) 
this function
goes to 1 quite fast. In this region the integral is cut off by the exponential
of momenta; so we can think of the exponential \( \exp \left\{ -\tau x(1-x)\frac{Q^{2}}{M^{2}_{c}}\right\}  \)
as providing a cutoff for \( \tau >4M^{2}_{c}/Q^{2} \), 
and \( F_{\delta }(\tau ) \)
as providing a cutoff for \( \tau <\pi ^{2} \). 
With this in mind, we can estimate \( \Pi _{\mathrm{fin}}(Q) \) as
\begin{equation}
\Pi _{\mathrm{fin}}(Q)\approx \frac{e_{4}^{2}}{2\pi ^{2}}\sum _{n}\int
^{1}_{0}dxx(1-x)\int ^{4M^{2}_{c}/Q^{2}}_{\pi ^{2}}\frac{d\tau }{\tau
}=-\frac{e^{2}}{2\pi ^{2}}\frac{1}{6}\log \frac{Q^{2}\pi ^{2}}{4M^{2}_{c}} ,\qquad Q^{2}<4M^{2}_{c}/\pi ^{2}
;\end{equation}
it is just the ordinary running of the zero mode. As \( Q^{2} \) grows,
the upper limit of integration is smaller than the lower limit,
and then we expect
that \( \Pi _{\mathrm{fin}}(Q) \) should vanish. 
\begin{figure}
\vspace{0.3cm}
{\par\centering \includegraphics{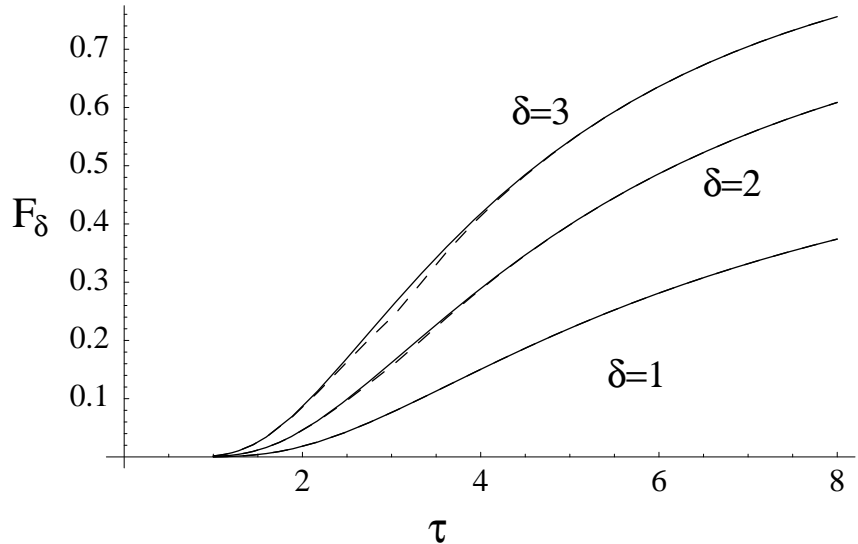} \par}
\vspace{0.3cm}
\caption{Exact values of \protect\( F_{\delta }(\tau )\protect \) (solid) as compared
with the approximation discussed in the text (dashed) for \protect\( \delta =1,2,3\protect \).\label{fig:approximation}}
\end{figure}
In that region \( \Pi (Q) \)
will be dominated completely by \( \Pi _{\mathrm{uc}}(Q) \).

Let us evaluate \( \Pi _{\mathrm{fin}}(Q) \) for any number 
of extra dimensions.
To this end we will approximate the function \( F_{\delta }(\tau ) \) as follows
\begin{equation}
\label{eq:approximation}
F_{\delta }(\tau )=\left\{ \begin{array}{ll}
2\delta \left( \frac{\pi }{\tau }\right) ^{\delta /2}\exp \left\{ -\frac{\pi ^{2}}{\tau }\right\} \quad \tau <\pi  & \\
1+2\delta \exp \left\{ -\tau \right\} -\left( \frac{\pi }{\tau }\right) ^{\delta /2}\quad \tau >\pi 
\end{array}\right. ~. 
\end{equation}
 The matching point in \( \tau =\pi  \) makes the function continuous. 
In Fig.~\ref{fig:approximation}
we display the exact function \( F_{\delta }(\tau ) \) (solid) and the approximation
above (dashed) for \( \delta =1,2,3 \).
The approximation is very good except at a small region around the matching
point \( \tau =\pi  \). This can be further improved by adding more terms from
the expansions of the \( \bar{\theta }(\tau ) \) functions.

The approximate expression of Eq.(\ref{eq:approximation}) can be used to 
obtain semi-analytical expansions for \( \Pi ^{\delta }_{\mathrm{fin}}(Q) \)
for small \( Q^{2} \) (we define \( w\equiv Q^{2}/M^{2}_{c} \)) 
\begin{eqnarray}
\Pi ^{(1)}_{\mathrm{fin}}(Q) & = & \frac{e_{4}^{2}}{2\pi ^{2}}
\left( -0.335-0.167\log (w)+0.463\sqrt{w}-0.110w+\cdots \right)
\label{eq:approx}~,\\
\Pi ^{(2)}_{\mathrm{fin}}(Q) & = & \frac{e_{4}^{2}}{2\pi ^{2}}
\left( -0.159-0.167\log (w)-0.105w\left( \log (w)-1.75\right) +\cdots
\right) ~, \\
\Pi ^{(3)}_{\mathrm{fin}}(Q) & = & \frac{e_{4}^{2}}{2\pi ^{2}}
\left( -0.0937-0.167\log (w)+0.298w-0.202\sqrt{w^{3}}+\cdots \right)~.
\end{eqnarray}

To see how good these approximate results are, we can compare with the exact
results that can be obtained when \( \delta =1 \). In this case we
have
\[
\Pi ^{(1)}_{\mathrm{fin}}(Q)=\frac{e_{4}^{2}}{2\pi ^{2}}\int ^{1}_{0}dxx(1-x)\sum ^{\infty }_{n=1}\int ^{\infty }_{0}\frac{d\tau }{\tau }\exp \left\{ -\tau x(1-x)w\right\} 2\left( \frac{\pi }{\tau }\right) ^{1/2}\exp \left\{ -\frac{n^{2}\pi ^{2}}{\tau }\right\} 
\]
\[
=\frac{e_{4}^{2}}{2\pi ^{2}}\int ^{1}_{0}dxx(1-x)\sum ^{\infty }_{n=1}\frac{2}{n}\exp \left\{ -2\pi n\sqrt{x(1-x)w}\right\} 
\]
\begin{equation}
\label{PI1full}
=\frac{e_{4}^{2}}{2\pi ^{2}}\int ^{1}_{0}dxx(1-x)\left( -2\log \left( 1-\exp
\left( -2\pi \sqrt{x(1-x)w}\right) \right) \right)~.
\end{equation}
This last expression 
can be expanded for \( w\ll 1 \), and the integral over \( x \) can then
be performed analytically, yielding 
\begin{eqnarray}
\Pi ^{(1)}_{\mathrm{fin}}(Q) &\approx& 
\frac{e_{4}^{2}}{2\pi ^{2}}\left( \frac{1}{18}
\left( 5-6\log (2\pi )\right) -\frac{1}{6}\log w+
\frac{3\pi ^{2}}{64}\sqrt{w}-\frac{\pi ^{2}}{90}w+\cdots \right)~, 
\end{eqnarray}
which is in excellent agreement with our approximation (see
Eq.~(\ref{eq:approx})).

Eq.~(\ref{PI1full}) can also be used to obtain the behavior of \( \Pi ^{(1)}_{\mathrm{fin}}(Q) \)
for \( w\gg 1. \) In this limit we obtain
\[
\Pi ^{(1)}_{\mathrm{fin}}(Q)\approx \frac{e_{4}^{2}}{2\pi ^{2}}\frac{3\zeta
(5)}{\pi ^{4}}\frac{M_{c}^{4}}{Q^{4}}\, ,\qquad Q^{2}\gg M^{2}_{c}~.
\]
 For higher dimensions things are more complicated, but the behavior is the
same, and we find 
\[
\Pi ^{(\delta )}_{\mathrm{fin}}(Q)\approx \frac{e_{4}^{2}}{2\pi
^{2}}\frac{4\delta \Gamma (2+\delta /2)}{\pi ^{4+\delta /2}}K_{\delta
}\frac{M^{4}_{c}}{Q^{4}}\, ,\qquad Q^{2}\gg M^{2}_{c}~,
\]
 where \( K_{\delta } \) is of the order of unity and is determined numerically
(\( K_{1}=\zeta (5)=1.037, \) \( K_{2}=1.165, \) \( K_{3}=1.244 \)). However,
since the uncompactified contribution grows as \( \left( Q^{2}/M^{2}_{c}\right) ^{\delta /2} \)
it is obvious that the contributions to \( \Pi ^{(\delta )}(Q) \) from \( \Pi ^{(\delta )}_{\mathrm{fin}}(Q) \)
will be completely irrelevant for \( Q^{2}\gg M^{2}_{c}. \)

Adding the finite and the uncompactified contributions we find that for \( Q^{2}\ll M^{2}_{c} \)
the uncompactified contribution exactly cancels the corresponding piece obtained
from the expansion of \( \Pi ^{(\delta )}_{\mathrm{fin}}(Q) \) (the \( \sqrt{w} \)
piece for \( \delta =1 \), the \( w\log (w) \) piece for \( \delta =2 \),
or the \( \sqrt{w^{3}} \) for \( \delta =3 \)). Then, for \( Q^{2}\ll M^{2}_{c} \)
and choosing \( \mu =M_{c} \) we finally find:

\begin{equation}
\label{PIfinal}
\Pi ^{(\delta )}(Q)=\frac{e_{4}^{2}}{2\pi ^{2}}\left( a^{(\delta )}_{0}-\frac{1}{6}\log \left( \frac{Q^{2}}{M^{2}_{c}}\right) +a^{(\delta )}_{1}\frac{Q^{2}}{M^{2}_{c}}+\cdots \right) \, ,\qquad Q^{2}\ll M^{2}_{c}
\end{equation}
 with the coefficients for 1,2 and 3 extra dimensions given by

\medskip{}
{\centering \begin{tabular}{|c|c|c|c|}
\hline 
\( \delta  \)&
 \( 1 \)&
 \( 2 \)&
 \( 3 \)\\
\hline 
\( a^{(\delta )}_{0} \)&
 \( -0.335 \)&
 \( -0.159 \)&
 \( -0.0937 \)\\
\hline 
\( a^{(\delta )}_{1} \)&
 \( -0.110 \)&
 \( 0.183 \)&
 \( 0.298 \) \\
\hline 
\end{tabular}\par}
\medskip{}

As we will see below, in general the coefficients \( a^{(\delta )}_{1} \) can
be affected by non-calculable contributions from higher dimension operators
in the effective Lagrangian, which we have not included.
 
The following comments related to Eq.~(\ref{PIfinal}), which is only valid
in the dimensional regularization scheme we are using,  are now in order:

({\bf i}) 
From Eq.~(\ref{PIfinal}) we see that for small \( Q^{2} \) , as expected,
we recover the standard logarithm with the correct coefficient, independently
of the number of extra dimensions. 
In addition, interestingly enough, we can compute
also the constant term. 
Thus, although the full theory in \( 4+\delta  \) dimensions
is non-renormalizable and highly divergent, the low energy limit 
of the VPF calculated in our dimensional regularization scheme
is actually finite: when seen from low energies the compactified extra 
dimensions seem to act as an ultraviolet regulator for the theory. 

({\bf ii})
When the energy begins to grow, we start
seeing effects suppressed by \( Q^{2}/M^{2}_{c} \),
which are finite, at one
loop, for any number of dimensions except for $\delta=2$. 
This is so because
the gauge couplings have dimensions \( 1/M^{\delta /2} \), and therefore, the
one-loop VPF goes like \( 1/M^{\delta } \). 

({\bf iii})
For \( \delta =1 \) one finds that,
because of gauge and Lorentz invariance, 
there are no possible counterterms
of this dimension. The VPF must be finite, and that is precisely the result
one obtains with dimensional regularization. This of course changes if higher
loops are considered: for instance, two-loop diagrams go like \( 1/M^{2} \),
and, in general, we expect that they will have divergences, which, in turn, 
should be absorbed in the appropriate counterterms. 
In principle the presence of these
counterterms could pollute our result; 
however, the natural size of these counterterms,
arising at two loops, should be suppressed compared to 
the finite contributions
we have computed. 

({\bf iv})
For \( \delta =2 \) one finds that the
VPF goes as \( 1/M^{2} \), already at one loop, and that the result is 
divergent. The divergences
have to be absorbed in the appropriate counterterm coming from higher dimension
operators in the higher dimensional theory. 
The immediate effect of this, is that 
the coefficient of the \( Q^{2} \) term in \( \Pi ^{(2)}(Q) \) 
becomes arbitrary, its value depending 
 on the underlying physics beyond
the compactification scale.

({\bf v}) For \( \delta >2 \) all loop contributions to the
\( Q^{2} \) term are finite, simply because of the dimensionality of the couplings.
This, however, does not preclude the 
existence of finite counterterms, which could
be generated by physics beyond the compactification scale, 
that is, contributions
from operators suppressed by two powers of the new physics scale like 
the operator in Eq.~(\ref{eq:lct}).

For \( Q^{2}\gg M^{2}_{c} \) the full VPF is completely 
dominated by the uncompactified contribution:

\begin{eqnarray*}
\Pi ^{(1)}(Q) & = & -\frac{e_{4}^{2}}{2\pi ^{2}}\frac{3\pi
^{2}}{64}\sqrt{\frac{Q^{2}}{M^{2}_{c}}}~,\\
\Pi ^{(2)}(Q) & = & \frac{e_{4}^{2}}{2\pi ^{2}}\frac{\pi
}{30}\frac{Q^{2}}{M^{2}_{c}}\log \left( \frac{Q^{2}}{M^{2}_{c}}\right) \,
\qquad Q^{2}\gg M^{2}_{c}~,\\
\Pi ^{(3)}(Q) & = & \frac{e_{4}^{2}}{2\pi ^{2}}\frac{5\pi ^{3}}{768}\left(
\frac{Q^{2}}{M^{2}_{c}}\right) ^{\frac{3}{2}}~.
\end{eqnarray*}

As before, the VPF could also receive
non-calculable contributions from higher dimension
operators which we have not included; in fact, for \( \delta =2 \), these
are needed to renormalize the VPF. How large can these non-calculable
contributions be?
Since our D-dimensional theory is an effective theory valid only for
$Q^2 \ll M_s^2$, even above \( M_{c} \) the results will be dominated by the
lowest power of $Q^2$. 
In the case of \( \delta =1 \), the first operator that one can write down
goes as \( Q^{2} \); therefore we expect that the one-loop contribution,
of order \( \sqrt{Q^{2}} \),
that we have computed, will dominate completely the result, as long as we do not 
stretch it beyond the applicability of the effective Lagrangian approach.
For \( \delta =2 \),
counterterms are certainly needed at order \( Q^{2} \); still one can hope that
the result will be dominated by the logarithm (as happens with chiral 
logarithms in \( \chi PT \)). 
For \( \delta =3 \) (and higher), the one loop result grows as 
\( \left(Q^{2}\right) ^{\delta/2} \);
however there could be operators giving contributions of  order \( Q^{2} \) with
unknown coefficients (in fact although in dimensional regularization those are
not needed, they must be included if cutoffs are used to regularize the theory).
Therefore,
unless for some reason they are absent from the theory, the result will be 
dominated
by those operators.

\section{Matching of gauge couplings\label{sec:matching}}

Using the VPF constructed in the previous section we can define 
a higher dimensional analogue of the conventional QED effective charge 
\cite{Itzykson:1980rh,Peskin:1995ev},
which will enter in any process involving off-shell photons, e.g.  
\begin{equation}
\frac{1}{\alpha _{\mathrm{eff}}(Q)}\equiv \frac{1}{\alpha _{4}}\left. \left( 1+\Pi ^{(\delta )}(Q)\right) \right| _{\msb _{\delta }}\, ,
\label{eq:DefinitionEffectiveCharge}
\end{equation}
where \( \alpha _{4}=e^{2}_{4}/(4\pi ) \). 
We remind the reader that \( e_{4} \)
denotes the (dimensionless) coupling of the four-dimensional theory including
all KK modes; it is directly related to the gauge coupling in the theory with
\( \delta  \) extra dimensions by Eq.(\ref{relcoup}). 
The subscript \( \msb _{\delta } \) means that the VPF has been regularized
using dimensional regularization in \( D=4+\delta -\epsilon  \) dimensions,
and that divergences, when present, are subtracted according to 
the \( \msb  \) procedure. 

To determine the relation between \( \alpha_{4} \) and the low energy coupling in
QED, we have to identify the effective charge computed in the compactified theory
with the low energy effective charge, at some low energy scale (for instance
\( Q^{2}=m^{2}_{Z}\ll M^{2}_{c} \)), where both theories are valid.
In that limit we can trust our approximate results of Eq.~(\ref{PIfinal}),
and write
\begin{equation}
\label{eq:matching}
\frac{1}{\alpha _{\mathrm{eff}}(m_{Z})}=\frac{1}{\alpha _{4}}+\frac{2}{\pi
}a^{(\delta )}_{0}-\frac{2}{3\pi }\log \left( \frac{m_{Z}}{M_{c}}\right)~.
\end{equation}
This equation connects the low energy QED coupling with the coupling in the
compactified D-dimensional theory, regularized by dimensional regularization. 
Note that this equation is completely independent of the way subtractions 
are performed
to remove the poles in \( 1/\epsilon  \). These poles only appear 
(and only for even number of dimensions) in the contributions proportional 
to \( Q^{\delta } \),
which vanish for \( Q\rightarrow 0 \). Eq.~(\ref{eq:matching}) contains,
apart from a finite constant, the standard logarithmic running from \( m_{Z} \)
to the compactification scale \( M_{c} \). 
It is interesting to notice that, 
in this approach, the logarithm
comes from the finite piece, and should therefore be considered as an infrared
(IR) logarithm. When seen from scales smaller than \( M_{c} \), these logarithms
appear to have an UV origin, while, when seen from scales above \( M_{c} \),
appear as having an IR nature.

It is important to emphasize that, in this scheme, 
the gauge coupling does not run any more above the compactification
scale. This seems counter-intuitive,
but it is precisely what happens in \( \chi PT \) when using dimensional regularization:
\( f_{\pi } \) does not run, it just renormalizes higher dimensional
operators~\cite{Weinberg:1979kz}. 

Now we can use Eq.~(\ref{eq:matching}) to write the effective charge at all
energies in terms of the coupling measured at low energies:
\begin{equation}
\label{eq:fullEffectiveCharge}
\frac{1}{\alpha _{\mathrm{eff}}(Q)}\equiv \frac{1}{\alpha
_{\mathrm{eff}}(m_{Z})}+\left. \frac{1}{\alpha _{4}}\left( \Pi ^{(\delta )}(Q)-\Pi ^{(\delta )}(m_{Z})\right) \right| _{\msb _{\delta }}~.
\end{equation}
 Note that the last term is independent of \( \alpha _{4} \) due to the implicit
dependence of \( \Pi ^{(\delta )} \) on it. Eq.~(\ref{eq:fullEffectiveCharge})
has the form of a momentum-subtracted definition of the coupling; in fact, in
four dimensions it is just the definition of the momentum-subtracted running
coupling. For \( \delta =1 \) and at one loop, 
\( \Pi ^{(\delta )}(Q)-\Pi ^{(\delta )}(m_{Z}) \)
is finite, and \( \alpha _{\mathrm{eff}}(Q) \) 
can still be interpreted as a momentum-subtracted 
definition of the coupling. 
For \( \delta >1 \), however, Eq.~(\ref{eq:fullEffectiveCharge})
involves additional subtractions, a fact which thwarts such an interpretation.

\begin{figure}
{\par\centering \includegraphics{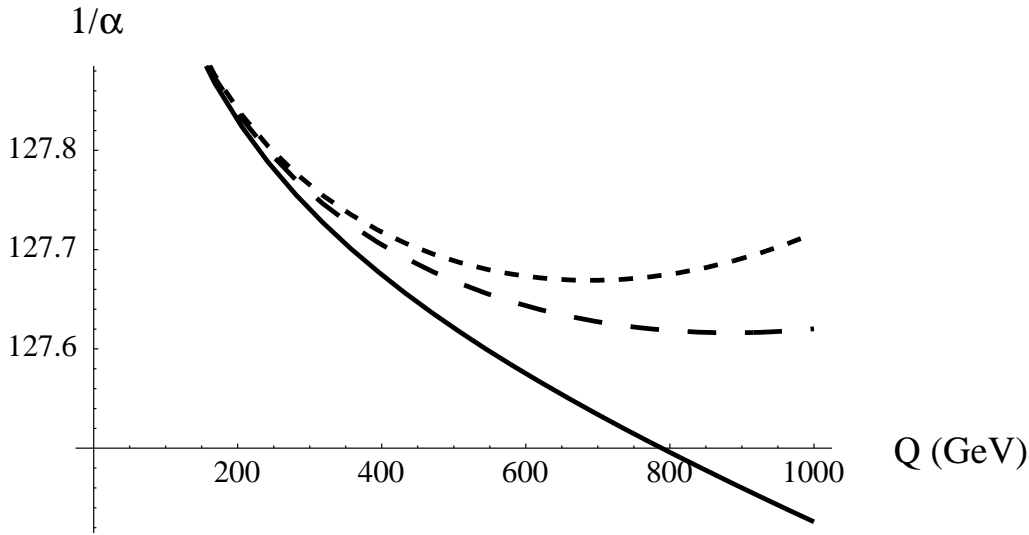} \par}
\caption{The {}``effective charge{}'' against the energy scale for \protect\( \delta =1\protect \)
(solid), \protect\( \delta =2\protect \) (short dash), \protect\( \delta =3\protect \)
(long dash).\label{fig:effective-running}}
\end{figure}

For \( Q^{2}\ll M^{2}_{c} \) we can expand \( \Pi ^{(\delta )}(Q) \) and obtain

\[
\frac{1}{\alpha _{\mathrm{eff}}(Q)}\equiv \frac{1}{\alpha
_{\mathrm{eff}}(m_{Z})}-\frac{2}{3\pi }\log \left( \frac{Q}{m_{Z}}\right)
+\mathcal{O}\left( \frac{Q^{2}}{M^{2}_{c}}\right)~, 
\]
which is nothing but the standard expression of the effective charge in QED,
slightly modified by small corrections of order \( Q^{2}/M^{2}_{c} \). However,
as soon as \( Q^{2}/M^{2}_{c} \) approaches unity, the effects of the 
compactification scale start to appear in \( \alpha _{\mathrm{eff}}(Q) \), 
forcing it to deviate dramatically from the logarithmic behavior, 
as shown in Fig.~\ref{fig:effective-running}.

The crucial point, however, is that this effective charge \emph{cannot} be interpreted
anymore as the running coupling (as can be done in four dimensions) since
it may receive contributions from higher dimension operators; in fact 
some of them are needed to define this quantity properly. These contributions have nothing 
to do with the gauge coupling which is defined as the coefficient of the operator 
\( F^{2}\). In particular, one should not use this quantity
to study gauge coupling unification. 
Instead, one could use Eq.~(\ref{eq:matching}), 
which relates the coupling measured at low energies with the one appearing in the 
\( D \)-dimensional Lagrangian valid at energies \( M_{c}<Q<M_{s} \). This
relation involves a logarithmic correction, which is the only contribution
that can be reliably computed without knowing the physics beyond \( M_{s} \).

It is instructive to see what happens if instead of dimensional
regularization we use hard cutoffs to regularize the uncompactified part of 
the VPF as in
Eqs.~(\ref{eq:PiCutoff})--(\ref{d3}). Then, when using cutoffs,
one can define an ``effective
charge'' as in Eq.~(\ref{eq:DefinitionEffectiveCharge})
\begin{equation}
\frac{1}{\alpha _{\mathrm{eff}}(Q)}\equiv \frac{1}{\alpha _{4}(\Lambda)}\left. \left( 1+\Pi ^{(\delta )}(Q)\right) 
\right| _{\Lambda}\, ,
\label{eq:DefinitionEffectiveChargeLambda}
\end{equation}
where $\alpha_4(\Lambda)$ now is the coupling constant in the theory
regularized with cutoffs and the subscript $\Lambda$ indicates that the VPF
has been regularized with cutoffs. The use of a $\Lambda$ dependent coupling
obviously implies the WEFT
formulation, in which the cutoff is not removed from the theory. 
On the other hand, in the CEFT
formulation one should renormalize the coupling constant by adding the
appropriate counterterms and then take the limit $\Lambda\rightarrow
\infty$. This usually brings in a new scale at which the coupling is
defined, and which effectively replaces $\Lambda$ in the previous equation.
Notice also that for $\delta=2$ in Eq.~(\ref{d2}) there are logarithmic
contributions proportional to $Q^2$, which cannot be removed when
$\Lambda\rightarrow \infty$. 
The same is true for $\delta>2$, but with
dependencies which are proportional to $\Lambda^{(\delta-2)}$. This just
manifests the need of higher dimensional operators, as was already clear in
the dispersive approach, to define properly the effective charge. 
As one can see, the full VPF
contains a term
that goes as \( \Lambda ^{\delta } \) and is independent of \( Q \). This
piece survives when \( Q\rightarrow 0 \), and thus  we obtain (we assume
$m_Z^2\ll Q^2$) 
\begin{equation}
\frac{1}{\alpha _{\mathrm{eff}}(m_{Z})}=
\frac{1}{\alpha _{4}(\Lambda )}+\frac{2}{3\pi \delta }
\left( \sqrt{\pi }\frac{\Lambda }{M_{c}}\right) ^{\delta }
+\frac{2}{\pi }a^{(\delta )}_{0}-\frac{2}{3\pi }
\log \left( \frac{m_{Z}}{M_{c}}\right) ~.
\end{equation}
Since $\alpha_{\mathrm{eff}}(m_Z)$ should be the same in the two schemes, 
we find the following relation between $\alpha_4$ and $\alpha_4(\Lambda)$
\begin{equation}
\frac{1}{\alpha _{4}}=\frac{1}{\alpha _{4}(\Lambda )}+\frac{2}{3\pi \delta
}\left( \sqrt{\pi }\frac{\Lambda }{M_{c}}\right) ^{\delta }~.
\label{match}
\end{equation}
If one identifies \(\Lambda  \) with the onset of a more complete theory
beyond the compactification scale, but at which the EFT treatment is still
valid, i.e. if one assumes that
\(\Lambda \sim M_G \ll M_{s} \), $M_G$ being this new scale,  Eq.(\ref{match})
could be reinterpreted as a matching equation
between the coupling \( \alpha _{4} \)  of our effective theory 
and the coupling
of the theory at scales \( M_G \), \( \alpha _{4}(M_G) \). 
Eq.~(\ref{match}) 
generically tells us that one expects 
corrections which go as \( \left( M_{G}/M_{c}\right) ^{\delta } \).
However, without knowledge of 
the full theory beyond \( M_{G} \), the meaning of
\( M_{G} \) (or even \( \alpha _{4}(M_{G}) \)) is unclear. In particular, if 
the new theory is some Grand Unified Theory in extra dimensions, 
\( M_{G} \) will be, in general, not just one single  
mass, but several
masses of the same order of magnitude, related by different coefficients. In
the case of logarithmic running those coefficients can be neglected, because
they give small logarithms 
next to the large logarithms containing the common scale. However,
in the case of contributions which depend on powers of the new physics scale
the situation is completely different, 
and the presence of several masses could change
completely the picture of unification. 
Cutoffs can give an indication of the
presence of power corrections, but the coefficients of these corrections cannot
be computed without knowing the details of the full theory. 

To see this point more clearly,
we add to our \( 4+\delta  \) dimensional theory an additional fermion 
with mass $M_f$ satisfying \( M_s\gg M_{f}\gg M_{c} \), 
such that  compactification corrections may be neglected,
and compute its effects on the coupling constant for 
\( M^2_{c}\ll Q^{2}\ll M^2_{f} \), using dimensional regularization. We have
\begin{equation}
\Pi ^{(\delta )}_{f}(Q)=\frac{e_{4}^{2}}{2\pi ^{2}}\left( \frac{\pi }{M_{c}}\right) ^{\delta /2}\Gamma (-\delta /2)\int ^{1}_{0}dxx(1-x)\left( M^{2}_{f}+x(1-x)Q^{2}\right) ^{\delta /2}\, .
\end{equation}
By expanding for \( Q^{2}\ll M_{f} \) and integrating over \( x \) 
we obtain
\begin{equation}
\Pi ^{(\delta )}_{f}(Q)=\frac{e_{4}^{2}}{2\pi ^{2}}\left( \sqrt{\pi }\frac{M_{f}}{M_{c}}\right) ^{\delta }\Gamma \left( -\frac{\delta }{2}\right) \left( \frac{1}{6}+\frac{\delta }{60}\frac{Q^{2}}{M_{f}^{2}}\right) \, .
\label{extraferm}
\end{equation}
 For odd values of \( \delta  \) we can use the analytic continuation of the
$\Gamma$ 
function to obtain a finite result. For even values of \( \delta  \)
we will allow a slight departure of the integer value in order to dimensionally
regularize the integral. 
Clearly, integrating out  
the heavy fermion  gives power corrections to the gauge
coupling. In addition, it 
also generates contributions 
to the higher dimension operators, e.g. 
contributions proportional \( Q^{2} \) and higher powers. 
As can be seen by comparing with Eqs.~(\ref{eq:PiCutoff})--(\ref{d3})
these power corrections are qualitatively similar 
to those calculated using a hard cutoff. 
Evidently, in the context of a more complete theory  
(in this case, given the existence of a heavy fermion), 
power corrections may be encountered 
even if the dimensional regularization is employed.
However, as one can easily see by setting, for example,  
$\delta = 1$ in Eq.(\ref{extraferm}),
the coefficients
of the power corrections obtained 
knowing the full theory 
are in general different from those 
obtained 
using a hard cutoff, 
e.g. Eq.(\ref{d1}). In fact, no choice of $\Lambda$ in 
Eq.(\ref{d1}) can reproduce all the coefficients appearing 
in Eq.(\ref{extraferm}). 

The situation is somewhat similar to what happens when $\chi PT$ with
$SU(2)\otimes SU(2)$ is matched to $\chi PT$ with $SU(3)\otimes SU(3)$.
In the $SU(2)\otimes SU(2)$ theory, just by dimensional arguments, one
can expect corrections like $m_K^2/f_\pi^2$. But, can one compute them
reliably without even knowing that there are kaons?

\section{Conclusions}

We have attempted a critical discussion of  
the arguments in favor of power-law running of coupling constants
in models with extra dimensions. We have shown that the naive arguments lead
to an arbitrary \( \beta  \) function depending on the particular way chosen
to cross KK thresholds. In particular, if one chooses 
the physical way of passing thresholds provided by the vacuum polarization 
function of the photon, a \( \beta  \)
function that counts the number of modes is divergent for more 
than \( 5 \) dimensions.

We have studied the question of decoupling of KK modes in QED with 4+ \( \delta  \)
(compact) dimensions by analyzing the behavior of the VPF of the photon. We
have computed first the imaginary part of the VPF by using unitarity
arguments,
and found that it rapidly reaches the value obtained in a non-compact theory
(only a few modes are necessary). We also showed that it grows as \(
(s/M^{2}_{c})^{\delta /2} \),
exhibiting clearly the non-renormalizability of theories in extra dimensions.
To obtain the full VPF, one can use an appropriately subtracted dispersion relation.
Instead, we use the full quantum effective field theory, with the expectation, 
suggested
by the calculation of the imaginary part of the VPF, that the bad UV behavior
of the theory is captured 
by the behavior of the uncompactified theory. To
check this idea, we have computed the VPF in the uncompactified theory,
regularized
by dimensional regularization (\( \delta \rightarrow \delta -\varepsilon  \)).
We have found that, after analytical continuation, the one loop VPF is finite,
and proportional to \( Q^{\delta } \) for odd number of dimensions, and has
a simple pole, proportional to \( Q^{\delta } \), for even number of dimensions.
This result can be understood easily, because there are no possible Lorentz and
gauge invariant operators in the 
Lagrangian able to absorb a term like \( Q^{\delta } \)
for odd \( \delta  \). For \( \delta  \) even it shows that higher dimension
operators are needed to regularize the theory. As a check we also recovered
the imaginary part of the VPF in the limit of 
infinite compactification radius.

For comparison with other approaches, we have also obtained the VPF in the case
that a hard cutoff is used to regularize it. We found that the pieces that do
not depend on the cutoff are exactly the same as those 
obtained by dimensional regularization,
while the cutoff dependent pieces are arbitrary, and can be changed at will by
changing the cutoff procedure.

Next we have computed the VPF in the compactified theory, and showed that it
can be separated 
into a UV and IR finite contribution and the VPF calculated in the 
uncompactified theory;  
as was shown previously, the latter 
can be controlled using dimensional regularization.
The finite part is more complicated, but can be computed numerically for any
number of dimensions. Also, some analytical approximations have been obtained
for the low and the high energy limits (\( Q\ll M_{c} \) and \( Q\gg M_{c} \)
respectively). Adding these two pieces, 
together with the counterterms coming from higher
dimension operators, we obtain 
a finite expression 
for an effective charge which can be extrapolated
continuously from \( Q\ll M_{c} \) to \( Q\gg M_{c} \); 
however, its value does depend 
on higher
dimension operator couplings.

Decoupling  of all KK  modes in  this effective  charge is  smooth and
physically  meaningful, and  the  low energy  logarithmic running  is
recovered.  We use  this effective  charge to  connect the  low energy
couplings (i.e. \( \alpha _{\mathrm{eff}}(m_{Z}) \)) with the coupling
of  the   theory  including  all   KK  modes,  regularized   by  dimensional
regularization. We find that  this matching only involves the standard
logarithmic running from \( m_{Z}  \) to the compactification scale \(
M_{c}  \).  In  particular, no  power corrections  appear  in this
matching.  However, if  cutoffs are used to regularize  the VPF in the
non-compact  space,  one  does  find  power  corrections,  exactly  as
expected from naive dimensional analysis.  In the EFT language one could
interpret  these corrections  as  an additional  matching between  the
effective  \( D  \) dimensional  field theory  and some  more complete
theory. The  question is how  reliably can this matching be estimated  
without knowing  the complete theory. By adding to our  theory 
an additional fermion with  \( M_{f}\gg  M_{c} \), 
and  integrating it  out, we
argue that  power corrections cannot  be computed without  knowing the
details of the complete theory, in which the \( D \) dimensional theory
is embedded. Some  examples in which this matching  can, in principle,
be        computed         are        some        string        models
\cite{Ghilencea:1998st,Kakushadze:1999bb,Choi:2002wx}, and the recently
proposed            de-constructed           extra           dimensions
\cite{Arkani-Hamed:2001ca,Cheng:2001vd,Pokorski:2001pv}.     For   the
question  of  unification  of   couplings  this  result  seems  rather
negative,  at   least  when  compared  with   standard  
Grand Unified Theories,
where gauge coupling unification can be tested without knowing 
their details.
Alternatively, one  can approach this  result from a  more optimistic
point of view,  and regard the requirement of  
low-energy unification  of couplings as a stringent constraint 
on  the possible completions
of              the             extra-dimensional             theories
\cite{Chankowski:2001hz,Arkani-Hamed:2001vr}.

\begin{acknowledgments}
This work has been funded by the Spanish MCyT under the Grants
BFM2002-00568 and FPA2002-00612, and by the OCyT of the
``Generalitat Valenciana'' under the Grant GV01-94.
\end{acknowledgments}


\end{document}